\documentclass[preprint,amsmath,amssymb,aps,showpacs]{revtex4}
\usepackage{graphicx}
\usepackage{bm}
\usepackage{subfigure}
\usepackage{amsmath}

\newcommand{\pr}{Phys. Rev.\ }

\newcommand{\etal}{{\em et al.}}

\newcommand{\jmp}{J. Math. Phys.\ }

\newcommand{\UQ}{School of Mathematics and Physics, University of Queensland, Brisbane, 
QLD 4072, Australia.}

\begin{document}

\title{Generalised purity as an entanglement measure for two- and three-well Bose-Hubbard models}

\author{C.~V. Chianca and M.~K. Olsen}
\affiliation{\UQ}
\date{\today}

\begin{abstract}

The $SU(2)$ and $SU(3)$ Lie algebras lend themselves naturally to studies of two- and three-well Bose-Einstein condensates, with the group operators being expressed in terms of bosonic annihilation and creation operators at each site. The success of these representations has led to the purities associated with these algebras to be promoted as a measure of entanglement between the atomic modes in each well. In this report, we show that these purities do not provide an unambiguous measure, but instead give results which depend on the quantum statistical states of each atomic mode. Using the example of totally uncoupled modes which have never interacted, we quantify these purities for different states and show that completely separable states can give values which have been claimed to indicate the presence of entanglement.  

\end{abstract}


\maketitle

\section{Introduction}
\label{sec:intro}

The $SU(2)$ and $SU(3)$ group operators used in the treatment of angular momentum and two- and three-well condensates originate from Schwinger's oscillator model of angular momentum~\cite{Julian}. In the case of a two-well condensate, and making the two-mode approximation, the relevance of Schwinger's model is apparent when we use the bosonic annihilation operators for each mode, $\hat{a}_{1}$ and $\hat{a}_{2}$ to construct three operators which obey SU(2) commutation relations~\cite{Joel,Marcos2},
\begin{eqnarray}
J_{x} &=& \frac{1}{2} (\hat{a}_{1}^{\dagger}\hat{a}_{1} - \hat{a}_{2}^{\dagger}\hat{a}_{2}),\nonumber \\ 
J_{y} &=& \frac{i}{2} (\hat{a}_{1}^{\dagger}\hat{a}_{2} - \hat{a}_{2}^{\dagger}\hat{a}_{1}), \nonumber \\
J_{z} &=& \frac{1}{2} (\hat{a}_{1}^{\dagger}\hat{a}_{2} + \hat{a}_{2}^{\dagger}\hat{a}_{1}).
\label{eq:JSU2}
\end{eqnarray}
We note here that we have used the operators as defined in Ref.~\cite{Marcos2}, in order to be consistent with the definition of the $SU(2)$ purity found in that article.
The most natural set of states which exhibit spontaneously broken symmmetry are then the coherent atomic states introduced by Arecchi \etal~\cite{Arecchi}, constructed from the Dicke states~\cite{Dicke}, which are themselves eigenstates of $J_{z}$.
These coherent atomic states exhibit a generalised SU(2) purity of one, which is the maximum value. 

In the case of a symmetric three-well condensate in the three-mode approximation~\cite{Viscondi}, it is natural to use operators based on the $SU(3)$ group generators,
\begin{eqnarray}
Q_{1} &=& \frac{1}{2} (\hat{a}_{1}^{\dagger}\hat{a}_{1} - \hat{a}_{2}^{\dagger}\hat{a}_{2}),  \qquad
Q_{2} = \frac{1}{3} (\hat{a}_{1}^{\dagger}\hat{a}_{1} + \hat{a}_{2}^{\dagger}\hat{a}_{2} - 2\hat{a}_{3}^{\dagger}\hat{a}_{3} ), \nonumber \\
J_{k} &=& i (\hat{a}_{k}^{\dagger}\hat{a}_{j} - \hat{a}_{j}^{\dagger}\hat{a}_{k}), \qquad
P_{k} = \hat{a}_{k}^{\dagger}\hat{a}_{j} + \hat{a}_{j}^{\dagger}\hat{a}_{k},
\label{eq:QJP}
\end{eqnarray}
where $k = 1,2,3$ and $j=(k+1)\text{mod3} + 1$. Note that $\hat{a}_{3}$ is the bosonic annihilation operator for the mode contained in the third well.  As with the two-mode system, atomic coherent states of the $SU(3)$ group may be defined~\cite{Mathur}, which are again the minimum uncertainty states of the relevant phase space and will therefore have an $SU(3)$ purity of one. 

Having defined the appropriate operators for each of these groups, we turn our attention to statements made that a generalised purity of less than one signifies entanglement in these bosonic systems. We will proceed by giving examples of particular quantum states in two- and three-well systems, and calculate the purities for these. Given that the Schwinger model was originally introduced for uncoupled oscillators, we are justified in defining initial quantum states in each well. We will not address whether or not it is possible to manufacture such states in the laboratory, but will use the fact that none of them can possibly be entangled, as they have never interacted and are completely separable. We will thus show that a generalised purity of less than unity is not a reliable signal of entanglement.  

\section{Generalised purity for the two-well model}
\label{sec:purity2} 

The generalised purity of the $SU(2)$ algebra is defined as~\cite{Marcos2} 
\begin{eqnarray}
\mathcal{P}_{SU(2)}(|\psi\rangle) = \frac{\langle J_{x} \rangle^2 + \langle J_{y} \rangle^2 + 
\langle J_{z} \rangle^2}{\langle J^2\rangle},
\label{eq:PSU2}
\end{eqnarray}
where $J^{2}=J_{x}^{2}+J_{y}^{2}+J_{x}^{2}$ and the expectation values are those for the state $|\psi\rangle$. It is a reasonably simple matter to evaluate this expression for a number of different quantum states. We will consider three different states for a system where the atoms in the two-modes have never interacted and thus cannot be entangled. These will be (i) an independent Glauber-Sudarshan coherent state in each well; (ii) an independent Fock state of fixed atom number in each well; and (iii) an independent coherently displaced squeezed state in each well~\cite{Danbook}.

\subsection{Coherent states}

We write a state with a coherent state in each well as $|\psi\rangle=|\alpha_{1},\alpha_{2}\rangle$, so that we have
$\hat{a}_{1}\hat{a}_{2}|\alpha_{1},\alpha_{2} \rangle = \alpha_{1}\alpha_{2} | \alpha_{1},\alpha_{2} \rangle$.  It is then a trivial matter to find the expectation values,
\begin{eqnarray}
\langle J_{x} \rangle &=& \frac{1}{2} (|\alpha_{1}|^{2} - |\alpha_{2}|^{2}), \nonumber \\ 
\langle J_{y} \rangle &=& \frac{i}{2} (\alpha_{1}^{\ast}\alpha_{2} - \alpha_{2}^{\ast}\alpha_{1}),\nonumber \\
\langle J_{z} \rangle &=& \frac{1}{2} (\alpha_{1}^{\ast}\alpha_{2} + \alpha_{2}^{\ast}\alpha_{1}).
\label{eq:J2coherent}
\end{eqnarray}
We can also calculate
\begin{eqnarray}
\langle J_{x}^{2}\rangle &=& \frac{1}{4}\left(|\alpha_{1}|^{4}+|\alpha_{2}|^{4}+|\alpha_{1}|^{2}+|\alpha_{2}|^{2}-2|\alpha_{1}|^{2}|\alpha_{2}|^{2}\right), \nonumber \\
\langle J_{y}^{2}\rangle &=& \frac{1}{4}\left(2|\alpha_{1}|^{2}|\alpha_{2}|^{2}+|\alpha_{1}|^{2}+|\alpha_{2}|^{2}-\alpha_{1}^{2}\alpha_{2}^{\ast\;2}-\alpha_{1}^{\ast\;2}\alpha_{2}^{2}\right), \nonumber\\
\langle J_{z}^{2}\rangle &=& \frac{1}{4}\left(2|\alpha_{1}|^{2}|\alpha_{2}|^{2}+|\alpha_{1}|^{2}+|\alpha_{2}|^{2}+\alpha_{1}^{\ast\;2}\alpha_{2}^{2}+\alpha_{1}^{2}\alpha_{2}^{\ast\;2}\right),
\label{eq:Jxyzsq}
\end{eqnarray}
so that 
\begin{equation}
\langle J^{2}\rangle = \frac{1}{4}\left(|\alpha_{1}|^{4}+|\alpha_{2}|^{4}+2|\alpha_{1}|^{2}|\alpha_{2}|^{2}+3|\alpha_{1}|^{2}+3|\alpha_{2}|^{2}\right),
\label{eq:JsqC}
\end{equation}
which is easily seen to equal the $\frac{N_{T}}{2}\left(\frac{N_{T}}{2}+1\right)$ given in ref.~\cite{Joel}, with $N_{T}$ being the expectation value of the total number of atoms.

It is now a trivial matter to calculate
\begin{equation}
\langle J_{x} \rangle^2 + \langle J_{y} \rangle^2 + \langle J_{z} \rangle^2 = \frac{1}{4}\left(|\alpha_{1}|^{2}+|\alpha_{2}|^{2}\right)^{2},
\label{eq:<J>sqC}
\end{equation}
so that the generalised purity for the Glauber-Sudarshan coherent states is
\begin{eqnarray}
\mathcal{P}_{SU(2)}(|\alpha_{1},\alpha_{2}\rangle) = \frac{\left(|\alpha_{1}|^{2}+|\alpha_{2}|^{2}\right)^{2}}{\left(|\alpha_{1}|^{2}+|\alpha_{2}|^{2}\right)^{2}+3\left(|\alpha_{1}|^{2}+|\alpha_{2}|^{2}\right)}.
\label{eq:PSU2C}
\end{eqnarray}
This expression is obviously always less than unity, and has an upper limit of unity in the limit of infinite coherent excitation.

\subsection{Fock states}

We now consider a system with an independent Fock state in each well, so that $|\psi\rangle=|n_{1},n_{2}\rangle$. It is immediately obvious that there is only one possible non-zero expectation value $\langle J_{k}\rangle$, which is
\begin{equation}
\langle J_{x}\rangle = \frac{1}{2}\left( n_{1}-n_{2}\right),
\label{eq:JxF}
\end{equation}
and we find the sum of the expectation values of the squares as
\begin{equation}
\langle J^{2}\rangle = \frac{1}{4}\left[(n_{1}+n_{2})^{2}+2(n_{1}+n_{2})\right].
\label{eq:J2F}
\end{equation}
This then gives the $SU(2)$ purity as
\begin{equation}
\mathcal{P}_{SU(2)}(|n_{1},n_{2}\rangle) = \frac{(n_{1}-n_{2})^{2}}{(n_{1}+n_{2})^{2}+2(n_{1}+n_{2})},
\label{eq:PSU2F}
\end{equation}
which can vary from zero when $n_{1}=n_{2}$ to a value which approaches unity when one of the wells has a much higher occupation than the other.

\subsection{Coherently displaced squeezed states}

For notational convenience we will write our squeezed states as $|s_{1},s_{2}\rangle$, where $s_{j}=\alpha_{j},r_{j}$, with $\alpha_{j}$ the coherent displacement and $r_{j}$ the squeezing parameter~\cite{Danbook}. Note that, in the interests of simplicity, we will consider $r_{j}$ and $\alpha_{j}$  to be real. Using the fact that such a state is produced by the action of first squeezing and then displacing the vacuum,
\begin{equation}
|s_{1},s_{2}\rangle = D(\alpha_{1})S(r_{1})D(\alpha_{2})S(r_{2})|0,0\rangle,
\label{eq:squeezemylemon}
\end{equation}
we may calculate all the required expectation values. We find
\begin{eqnarray}
\langle J_{x}\rangle &=& \frac{1}{2}\left(\alpha_{1}^{2}+\sinh^{2}r_{1}-\alpha_{2}^{2}-\sinh^{2}r_{2}\right),\nonumber \\
\langle J_{y}\rangle &=& \frac{i}{2}\left(\alpha_{1}\alpha_{2}-\alpha_{1}\alpha_{2}\right),\nonumber \\
\langle J_{z}\rangle &=& \frac{1}{2}\left(\alpha_{1}\alpha_{2}+\alpha_{1}\alpha_{2}\right),
\label{eq:Jzyzsqueezed}
\end{eqnarray}
so that
\begin{eqnarray}
\langle J_{x}\rangle^{2}+\langle J_{y}\rangle^{2}+\langle J_{z}\rangle^{2} &=& \frac{1}{4}\left[\left(\alpha_{1}^{2}+\alpha_{2}^{2}\right)^{2}+2\left(\alpha_{1}^{2}-\alpha_{2}^{2}\right)\left(\sinh^{2}r_{1}-\sinh^{2}r_{2}\right)\right.  \nonumber \\
& & \left. +\left(\sinh^{2}r_{1}-\sinh^{2}r_{2}\right)^{2}\right],
\label{eq:<J>sqsqueeze}
\end{eqnarray}
which we can see is the same expression as for a coherent state when $r_{j}=0$.
We now turn to the terms in the demoninator, finding
\begin{eqnarray}
\langle J_{x}^{2}\rangle &= & \frac{1}{4}\left[
\left(\alpha_{1}^{2}-\alpha_{2}^{2}\right)^{2}+\alpha_{1}^{2}\left(\cosh r_{1}-\sinh r_{1}\right)^{2} +\alpha_{2}^{2}\left(\cosh r_{2}-\sinh r_{2}\right)^{2}\right.\nonumber \\
& & \left.
+\left(\sinh^{2}r_{1}-\sinh^{2}r_{2}\right)^{2}+2\left(\sinh^{2}r_{1}\cosh^{2}r_{1}+\sinh^{2}r_{2}\cosh^{2}r_{2}\right)\right], \nonumber \\
\langle J_{y}^{2}\rangle &=& \frac{1}{4}\left[\alpha_{1}^{2}\left(\sinh r_{2}+\cosh r_{2}\right)^{2}+\alpha_{2}^{2}\left(\sinh r_{1}+\cosh r_{1}\right)^{2}+\left(\sinh r_{1}\cosh r_{2}-\sinh r_{2}\cosh r_{1}\right)^{2}
\right], \nonumber \\
\langle J_{z}^{2}\rangle &=& \frac{1}{4}\left[4\alpha_{1}^{2}\alpha_{2}^{2}+\alpha_{1}^{2}\left(\sinh r_{2}-\cosh r_{2}\right)^{2}+\alpha_{2}^{2}\left(\sinh r_{1}-\cosh r_{1}\right)^{2}\right. \nonumber \\
& & \left.
+\left(\sinh r_{1}\cosh r_{2}+\sinh r_{2}\cosh r_{1}\right)^{2}\right],
\label{eq:Jsqcomprimidos2}
\end{eqnarray}
so that 
\begin{eqnarray}
\langle J^{2}\rangle &=& \frac{1}{4}\left\{\left(\alpha_{1}^{2}+\alpha_{2}^{2}\right)^{2}+\alpha_{1}^{2}\left[\left(\cosh r_{1}-\sinh r_{1}\right)^{2}+2\left(\sinh^{2}r_{2}+\cosh^{2}r_{2}\right)\right]\right. \nonumber\\
& &\left.
+\alpha_{2}^{2}\left[\left(\cosh r_{2}-\sinh r_{2}\right)^{2}+2\left(\sinh^{2}r_{1}+\cosh^{2}r_{1}\right)\right]\right.\nonumber\\
& &\left.
+2\left(\sinh^{2}r_{1}+\sinh^{2}r_{2}\right)\left(\cosh^{2}r_{1}+\cosh^{2}r_{2}\right)
\right\}.
\label{eq:Jsumsq2}
\end{eqnarray}
The expression for $P_{SU(2)}|s_{1},s_{2}>$ is therefore rather complicated and large, but we can evaluate it readily for some special cases. Firstly, when $r_{1}=r_{2}=0$, so that we have two independent Glauber-Sudarshan coherent states, we find the same result as that given above in Eq.~\ref{eq:PSU2C}, as required. For two squeezed states with zero coherent excitation, we find
\begin{equation}
\mathcal{P}_{SU(2)}(|r_{1},r_{2}\rangle) = \frac{\left(\sinh^{2}r_{1}-\sinh^{2}r_{2}\right)^{2}}{2\left(\sinh^{2}r_{1}+\sinh^{2}r_{2}\right)\left(\cosh^{2}r_{1}+\cosh^{2}r_{2}\right)},
\label{eq:PSU2squeeze}
\end{equation}
which is zero if $r_{1}=r_{2}$ and tends towards one half for $r_{1}\gg r_{2}$.

\section{Generalised purity for the three-well model}
\label{sec:PSU3}

We will now consider the generalised purity associated with the $SU(3)$ algebra which is defined by Viscondi \etal~\cite{Viscondi} as 
\begin{eqnarray}
\mathcal{P}_{SU(3)}(|\psi\rangle) = \frac {9}{\langle N^2\rangle} \left( \frac {\langle \psi| Q_{1}|\psi\rangle^2}{3} + \frac {\langle \psi| Q_{2}|\psi\rangle^2}{4} + \sum_{j=1}^{3} \frac {\langle \psi| P_{j}|\psi\rangle^2}{12} + \sum_{k=1}^{3} \frac {\langle \psi| J_{k}|\psi\rangle^2}{12}\right),
\label{eq:PSU3def}
\end{eqnarray}
where $P_{i},\:J_{i}$ and $Q_{1}$ are as defined in the introduction, Eq.~\ref{eq:QJP}, and $N=\sum_{i=1}^{3}\hat{a}_{i}^{\dag}\hat{a}_{i}$. It has been stated in various publications that states with $\mathcal{P}_{SU(3)}(|\psi\rangle) = 1$ are separable, with any decrease from this maximum value indicating entanglement among the particles~\cite{Viscondi,Viola1,Viola2}. We will now evaluate this purity for the three-mode analogues of the separable states considered in section~\ref{sec:purity2}. 

\subsection{Independent coherent states}
\label{sec:tricoherent}

We consider independent occupations of each well by Glauber-Sudarshan coherent states, $|\psi\rangle=|\alpha_{1},\alpha_{2},\alpha_{3}\rangle$, and calculate
\begin{eqnarray}
\langle Q_{1} \rangle^2 &=& \frac {1}{4} \left( |\alpha_{1}|^{2}  - |\alpha_{2}|^{2} \right)^{2}, \nonumber \\
\langle Q_{2} \rangle^2 &=& \frac {1}{9} \left( |\alpha_{1}|^{2} + |\alpha_{2}|^{2} - 2|\alpha_{3}|^{2} \right)^{2}, \nonumber \\
\langle P_{1} \rangle^2 &=& \left(\alpha_{1}^{\ast}\alpha_{3} +\alpha_{3}^{\ast}\alpha_{1}\right)^{2}, \nonumber \\
\langle P_{2} \rangle^2 &=& \left(\alpha_{2}^{\ast}\alpha_{1}+ \alpha_{1}^{\ast}\alpha_{2}\right)^{2}, \nonumber \\
\langle P_{3} \rangle^2 &=& \left(\alpha_{3}^{\ast}\alpha_{2} +  \alpha_{2}^{\ast}\alpha_{3}\right)^{2}, \nonumber \\
\langle J_{1} \rangle^2 &=& -\left(\alpha_{1}^{\ast}\alpha_{3} - \alpha_{3}^{\ast}\alpha_{1}\right)^{2}, \nonumber \\
\langle J_{2} \rangle^2 &=& -\left(\alpha_{2}^{\ast}\alpha_{1} - \alpha_{1}^{\ast}\alpha_{2}\right)^{2}, \nonumber \\
\langle J_{3} \rangle^2 &=& -\left(\alpha_{3}^{\ast}\alpha_{2} - \alpha_{2}^{\ast}\alpha_{3}\right)^{2},
\label{eq:QPK}
\end{eqnarray}
as well as
\begin{eqnarray}
\langle N^{2}\rangle &=& \left(|\alpha_{1}|^{2}+|\alpha_{2}|^{2}+|\alpha_{3}|^{2}\right)^{2}+|\alpha_{1}|^{2}+|\alpha_{2}|^{2}+|\alpha_{3}|^{2} \nonumber \\
&=& \langle N\rangle^{2}+\langle N\rangle.
\label{eq:alpha3sq}
\end{eqnarray}
It is the a simple matter to combine these expressions as in Eq.~\ref{eq:PSU3def} to find
\begin{eqnarray}
\mathcal{P}_{SU(3)}(|\alpha_{1},\alpha_{2},\alpha_{3}\rangle) &=& \frac{\left(|\alpha_{1}|^{2}+|\alpha_{2}|^{2}+|\alpha_{3}|^{2}\right)^{2}}{\left(|\alpha_{1}|^{2}+|\alpha_{2}|^{2}+|\alpha_{3}|^{2}\right)^{2}+|\alpha_{1}|^{2}+|\alpha_{2}|^{2}+|\alpha_{3}|^{2}} \nonumber \\
&=& \frac{\langle N\rangle^{2}}{\langle N\rangle^{2}+\langle N\rangle}.
\label{eq:PSU3Roy}
\end{eqnarray}
It is readily seen that this value will always be smaller than one, approaching one in the limit of extremely large $N$.

\subsection{Independent Fock States}
\label{sec:Fock3}

We now turn our attention to three independent Fock states, $|n_{1},n_{2},n_{3}\rangle$. We find
\begin{eqnarray}
Q_{1} &=& \frac{1}{2}\left(n_{1}-n_{2}\right), \nonumber \\
Q_{2} &=& \frac{1}{3}\left(n_{1}+n_{2}-2n_{3}\right),\nonumber \\
J_{k} &=& P_{k} =0.
\label{eq:PQJFock3}
\end{eqnarray}
After a little simple algebra, we find
\begin{equation}
\mathcal{P}_{SU(3)}(|n_{1},n_{2},n_{3}\rangle) 
= 1 - \frac {3(n_{1}n_{2} + n_{1}n_{3} + n_{2}n_{3})}{(n_{1} + n_{2} + n_{3})^{2}}.
\label{eq:PSU3Fock}
\end{equation}
It is readily seen that this will be equal to zero when $n_{1}=n_{2}=n_{3}$ and can take on a range of values when one well is much more highly occupied than the others.

\subsection{Independent Squeezed States}
\label{sec:3squeeze}

In the case of three independent coherently displaced squeezed states, with $\alpha_{j}$ the coherent displacements and $r_{j}$ the squeezing parameters, we find
\begin{equation}
\langle Q_{1} \rangle^2 = \frac {1}{4} \left(|\alpha_{1}|^{2}+\sinh^{2}r_{1}- |\alpha_{2}|^{2}-\sinh^{2}r_{2}\right)^{2},
\label{eq:Q1sq3}
\end{equation}
and
\begin{eqnarray}
\langle Q_{2} \rangle^2 &=& \frac {1}{9} \left[ |\alpha_{1}|^{2}+\sinh^{2}r_{1}+|\alpha_{2}|^{2}+\sinh^{2}r_{2}-2\left(|\alpha_{3}|^{2}+\sinh^{2}r_{3}\right)\right]^{2},
\label{eq:Q2sq3}
\end{eqnarray}
with the $P_{k}$ and $J_{k}$ being the same as for coherent states, see Eq.~\ref{eq:QPK}. This gives the numerator as
\begin{eqnarray}
\mathcal{N} &=& \left(|\alpha_{1}|^{2}+\sinh^{2}r_{1}+|\alpha_{2}|^{2}+\sinh^{2}r_{2}+|\alpha_{3}|^{2}+\sinh^{2}r_{3}\right)^{2}-3\left[|\alpha_{1}|^{2}\left(\sinh^{2}r_{2}+\sinh^{2}r_{3}\right)\right.\nonumber\\
& &\left.
+|\alpha_{2}|^{2}\left(\sinh^{2}r_{1}+\sinh^{2}r_{3}\right)+|\alpha_{3}|^{2}\left(\sinh^{2}r_{1}+\sinh^{2}r_{2}\right)\right].
\label{eq:PSU3sqnum}
\end{eqnarray}
The denominator is found as
\begin{eqnarray}
\langle N^{2}\rangle &=& \left(|\alpha_{1}|^{2}+\sinh^{2}r_{1}+|\alpha_{2}|^{2}+\sinh^{2}r_{2}+|\alpha_{3}|^{2}+\sinh^{2}r_{3}\right)^{2}\nonumber\\
& & +|\alpha_{1}|^{2}\cosh^{2}r_{1}+|\alpha_{2}|^{2}\cosh^{2}r_{2}+|\alpha_{1}|^{3}\cosh^{2}r_{3}\nonumber\\
& & + 2\left(\sinh^{2}r_{1}\cosh^{2}r_{1}+\sinh^{2}r_{2}\cosh^{2}r_{2}+\sinh^{2}r_{3}\cosh^{2}r_{3}\right).
\label{eq:N3sq}
\end{eqnarray}
Again we see that the full expression for $\mathcal{P}_{SU(3)}$ is complicated, but easy to evaluate in some special cases. For example, when $r_{j}=0$, we find the same value as for coherent states, given in Eq.~\ref{eq:PSU3Roy}. When the $\alpha_{j}$ are all set to zero, we find
\begin{equation}
\mathcal{P}_{SU(3)}(|r_{1},r_{2},r_{3}\rangle) = \frac{N_{T}^{2}}{N_{T}^{2}+2\left(\sinh^{2}r_{1}\cosh^{2}r_{1}+\sinh^{2}r_{2}\cosh^{2}r_{2}+\sinh^{2}r_{3}\cosh^{2}r_{3}\right)},
\label{eq:PSU3alpha0}
\end{equation}
where $N_{T}=\sinh^{2}r_{1}+\sinh^{2}r_{2}+\sinh^{2}r_{3}$. It is readily seen that, whatever combination of $\alpha$ and $r$ we choose, the purity will be less than one, despite the fact that the states have been constructed so as to be completely separable.

\section{Conclusions}

In conclusion, we have shown that the generalised $SU(2)$ and $SU(3)$ purities are not a valid entanglement measure for multi-mode continuous variable systems by considering the cases of two- and three-well Bose-Hubbard models and demonstrating that fully separable states can be constructed which give  a value of less than one. This suggests strongly that great care should be used with this measure if it is desired to use it as a signature of quantum entanglement, and that by itself it is not sufficient. What it does measure is the "distance" of a quantum state from one of the $SU(N)$ coherent states, which is not necessarily related to entanglement in any way.

\section*{Acknowledgments}

This research was supported by the Australian Research Council under the Future Fellowships scheme. We wish to thank Simon Haine for help with operator calculations in Mathematica.


\begin{thebibliography}{99}
%
\bibitem{Julian}{J. Schwinger, in {\em Quantum Theory of Angular Momentum},
eds. L. C. Biedenharn and H. van Dam (Academic, New York, 1965.)}
%
\bibitem{Joel}{G.J. Milburn, J. Corney, E.M. Wright, and D.F. Walls, \pra {\bf 55}, 4318 (1997).}
%
\bibitem{Marcos2}{T.F. Viscondi, K. Furuya, and M.C. de Oliveira, \pra {\bf 80}, 013610 (2009).}
%
\bibitem{Arecchi}{F.T. Arrechi, E. Courtens, R. Gilmore, and H. Thomas, \pra {\bf 6}, 2211 (1972).}
%
\bibitem{Dicke}{R.H. Dicke, \pr {\bf 93}, 99 (1954).}
%
\bibitem{Viscondi}{T.F. Viscondi, K. Furuya, and M.C. de Oliveira, EPL {\bf 90}, 10014 (2010).}
%
\bibitem{Mathur}{M. Mathur and D. Sen, \jmp {\bf 42}, 4181 (2001).}
%
\bibitem{Danbook}{D.F. Walls and G.J. Milburn, {\em Quantum Optics}, (Springer-Verlag, Berlin, 1994).}
%
\bibitem{Viola1}{H. Barnum, E. Knill, G. Ortiz, and L. Viola, \pra {\bf 68}, 032308 (2003).}
%
\bibitem{Viola2}{R. Somma, G. Ortiz, H. Barnum, E. Knill, and L. Viola, \pra {\bf 70}, 042311 (2004).}
%

%
\end{thebibliography}
\end{document}